\input harvmac
\noblackbox
%


\font\cmss=cmss10
\font\cmsss=cmss10 at 7pt
\def\rlx{\relax\leavevmode}
\def\inbar{\vrule height1.5ex width.4pt depth0pt}
\def\IC{\relax\,\hbox{$\inbar\kern-.3em{\rm C}$}}
\def\IN{\relax{\rm I\kern-.18em N}}
\def\IP{\relax{\rm I\kern-.18em P}}
\def\ZZ{\rlx\leavevmode\ifmmode\mathchoice{\hbox{\cmss Z\kern-.4em Z}}
 {\hbox{\cmss Z\kern-.4em Z}}{\lower.9pt\hbox{\cmsss Z\kern-.36em Z}}
 {\lower1.2pt\hbox{\cmsss Z\kern-.36em Z}}\else{\cmss Z\kern-.4em
 Z}\fi}
\def\IZ{\relax\ifmmode\mathchoice
{\hbox{\cmss Z\kern-.4em Z}}{\hbox{\cmss Z\kern-.4em Z}}
{\lower.9pt\hbox{\cmsss Z\kern-.4em Z}}
{\lower1.2pt\hbox{\cmsss Z\kern-.4em Z}}\else{\cmss Z\kern-.4em
Z}\fi}

\def\narrowplus{\kern -.04truein + \kern -.03truein}
\def\narrowminus{- \kern -.04truein}
\def\narrowminussub{\kern -.02truein - \kern -.01truein}

\def\frac#1#2{{#1\over #2}}

\def\IZ{\relax\ifmmode\mathchoice
{\hbox{\cmss Z\kern-.4em Z}}{\hbox{\cmss Z\kern-.4em Z}}
{\lower.9pt\hbox{\cmsss Z\kern-.4em Z}}
{\lower1.2pt\hbox{\cmsss Z\kern-.4em Z}}\else{\cmss Z\kern-.4em
Z}\fi}
\def\IB{\relax{\rm I\kern-.18em B}}
\def\IC{{\relax\hbox{$\inbar\kern-.3em{\rm C}$}}}
\def\ID{\relax{\rm I\kern-.18em D}}
\def\IE{\relax{\rm I\kern-.18em E}}
\def\IF{\relax{\rm I\kern-.18em F}}
\def\IG{\relax\hbox{$\inbar\kern-.3em{\rm G}$}}
\def\IGa{\relax\hbox{${\rm I}\kern-.18em\Gamma$}}
\def\IH{\relax{\rm I\kern-.18em H}}
\def\II{\relax{\rm I\kern-.18em I}}
\def\IK{\relax{\rm I\kern-.18em K}}
\def\IP{\relax{\rm I\kern-.18em P}}

\font\cmss=cmss10 \font\cmsss=cmss10 at 7pt
\def\IR{\relax{\rm I\kern-.18em R}}

\def\1{{\bf 1}}
\def\3{{\bf 3}}
\def\7{{\bf 7}}
\def\6{{\bf 6}}
\def\2{{\bf 2}}
\def\8{{\bf 8}}

\def\muu#1{{\mu^{(u)}_#1}}

\lref\bansuss{T. Banks and L. Susskind, {\it ``Brane--Antibrane
Forces''}, hep-th/9511194.}
\lref\sred{M. Srednicki, {\it ``IIB or Not IIB''}, hep-th/9807138;
JHEP {\bf 08} (1998) 005.}
\lref\senreview{A. Sen, {\it ``Non-BPS States and Branes in String
Theory''}, hep-th/9904207.}
\lref\senworld{A. Sen, {\it ``Supersymmetric World-Volume Action For
Non-BPS D-Branes''}, hep-th/9909062; JHEP {\bf 10} (1999) 008.} 
\lref\sentach{A. Sen, {\it ``Tachyon Condensation on the Brane
Anti-Brane System''}, hep-th/9805170; JHEP {\bf 08} (1998) 012.}
\lref\senuniv{A. Sen, {\it ``Universality of the Tachyon Potential''}, 
hep-th/9911116.}
\lref\senbound{A. Sen, {\it ``Stable Non-BPS Bound States of 
BPS D-branes''}, hep-th/9805019; JHEP {\bf 08} (1998) 010.}
\lref\sencycle{A. Sen, {\it ``BPS D-Branes on Nonsupersymmetric
Cycles''}, hep-th/9812031; JHEP {\bf 12} (1998) 021.}
\lref\joysen{J. Majumder and A. Sen, {\it `` `Blowing Up' D-Branes 
on Nonsupersymmetric Cycles''}, hep-th/9906109; JHEP {\bf 09} (1999)
004.} 
\lref\witk{E. Witten, {\it ``D-branes and K-theory''},
hep-th/9810188; JHEP {\bf 12} (1998) 019.}
\lref\hork{P. Horava, {\it ``Type IIA D-Branes, K-Theory and Matrix
Theory''}, hep-th/9812135; Adv. Theor. Math. Phys. {\bf 2} (1999)
1373.}
\lref\bergabrev{O. Bergman and M. Gaberdiel, {\it ``Non-BPS Dirichlet
Branes''}, hep-th/9908126.}
\lref\tatar{M. Mihailescu, K. Oh and R. Tatar, {\it ``Non-BPS Branes 
on a Calabi-Yau Threefold and Bose-Fermi Degeneracy''}, hep-th/9910249.}
\lref\lerda{A. Lerda and R. Russo, {\it ``Stable Non-BPS States in 
String Theory: A Pedagogical Review''}, hep-th/9905006.} 
\lref\seiwit{N. Seiberg and E. Witten, {\it ``String Theory and
Noncommutative Geometry''}, hep-th/9908142; JHEP {\bf 09} (1999) 032.}
\lref\gms{R. Gopakumar, S. Minwalla and A. Strominger, {\it ``Noncommutative
Solitons''}, hep-th/0003160. }
\lref\bcr{M. Bill\`o, B. Craps and F. Roose, {\it ``Ramond-Ramond Coupling
of Non-BPS D-Branes''}, hep-th/9905157; JHEP {\bf 06} (1999) 033. }
\lref\wati{W. Taylor, {\it ``D2 Branes in B fields''}, hep-th/0004141.}
\lref\mst{ S. Mukhi, N. V. Suryanarayana and D. Tong, {\it ``Brane-
Antibrane Constructions''}, hep-th/0001066; JHEP {\bf 03} (2000) 015.}
\lref\panda{E. A. Bergshoeff, M. de Roo, T. C. de Wit, E. Eyras and S. Panda,
 {\it ``T-Duality and Actions for Non-BPS D-Branes''}, hep-th/0003221.}
\lref\garou{M. R. Garousi, {\it ``Tachyon Coupling on Non-BPS D-Branes 
and Dirac-Born-Infeld Action''}, hep-th/0003122.}
\lref\berggab{O. Bergman and M. Gaberdiel, {\it ``Stable Non-BPS 
D-Particle''}, hep-th/9806155; Phys. Lett. {\bf B441} (1998) 133.}
\lref\senso{A. Sen, {\it ``SO(32) Spinors of Type I and other Brane-
Antibrane Pair''}, hep-th/9808141; JHEP {\bf 09} (1998) 023.}
\lref\sentypi{A. Sen, {\it ``Type I D-Particle and its Interactions''},
hep-th/9809111; JHEP {\bf 10} (1998) 021.}
\lref\harvey{J.~Harvey, P.~Kraus, F. Larsen, E.~Martinec, to appear.}

\Title{\vbox{\hbox{hep-th/0005006}
\hbox{IASSNS-HEP-00/36}
\hbox{TIFR/TH/00-20}}}
{\vbox{\centerline{Noncommutative Tachyons}}}
\centerline{Keshav Dasgupta$^{a,}$\footnote{$^1$}{keshav@sns.ias.edu}, 
Sunil Mukhi$^{b,}$\footnote{$^2$}{mukhi@tifr.res.in} and
Govindan Rajesh$^{a,}$\footnote{$^3$}{rajesh@sns.ias.edu}}
\vskip 0.1 in
\medskip\centerline{$^a$\it School of Natural Sciences}
\centerline{\it Institute for Advanced Study}\centerline{\it
Princeton, NJ
08540, USA}
\medskip
\centerline{$^b$\it Tata Institute of Fundamental Research}
\centerline{\it Homi Bhabha Road, Mumbai 400 005, India}

\vskip 0.5in

When unstable Dp-branes in type II string theory are placed in a
B-field, the resulting tachyonic world-volume theory becomes
noncommutative. We argue that for large noncommutativity parameter,
condensation of the tachyon as a noncommutative soliton leads to new
decay modes of the Dp-brane into (p-2)-brane configurations, which we
interpret as suitably smeared BPS D(p-1)-branes. Some of these
configurations are metastable. We discuss various generalizations of
this decay process.

\Date{4/00}

\newsec{Introduction}

The decay modes of unstable D-branes and brane-antibrane pairs have
been extensively studied in the last couple of years (for reviews see
for example Refs.\refs{\senreview,\lerda,\bergabrev}).  Some of the
most basic decay modes found so far are: annihilation into
vacuum\refs{\bansuss,\sred,\senbound,\sentach}, annihilation via kink
condensation into a brane of codimension
one\refs{\berggab,\senso,\sentypi,\hork}, and annihilation via vortex
condensation to a brane of codimension
two\refs{\senso,\witk}. Condensation of higher-codimension topological
solitons has also been studied\refs{\witk,\hork}.  Some of these decay
modes correspond to stable solitons, and in this case the end-products
are stable branes, while in other cases the decay modes correspond to
unstable solitons and lead to unstable branes. More exotic decay modes
are possible when unstable branes or brane-antibrane pairs are either
wrapped on homology cycles of nontrivial
manifolds\refs{\sencycle,\joysen,\tatar}\ or suspended between other
branes in brane constructions\mst.

In this note, we identify a class of novel processes in which unstable
D-branes decay via condensation of a noncommutative soliton. Such
solitons\gms\ are finite-energy classical solutions in noncommutative
scalar field theories with large noncommutativity parameter
$\theta_{ij}$. Physically, this situation can be realised by turning
on a suitable constant NS-NS $B$-field along the brane world-volume.

Since noncommutative solitons carry no topological charge and are
sometimes metastable, the same will be true of the decay products of
an unstable brane when such a soliton condenses on its worldvolume. We
will interpret these decay products by examining what Ramond-Ramond
charge they carry locally.

While this work was nearing completion, we became aware of a
forthcoming paper\harvey\ which addresses related questions, mainly in
the context of the bosonic string.

\newsec{Noncommutative Tachyons and Brane Decay}

Consider an infinitely extended, unstable D2-brane in type IIB string
theory. We place it in a $B_{NS,NS}$ field with a nonzero component
only along the brane world-volume directions, namely $B_{1\,2}$.
Consider the brane world-volume theory. The light modes are a real
tachyon, a massless U(1) gauge field and some massless fermions.

Let us set all fields on the brane, except the tachyon, to 0. The
spacetime backgrounds are also as simple as possible: a flat metric
and the constant $B_{1\,2}$ field.  Following arguments in
Ref.\seiwit, the effect of the $B$-field can be described by making
the tachyonic field theory on the D2-brane world-volume into a
noncommutative field theory with $*$ product given by
\eqn\starprod{
f(x)*g(x) = e^{{i\over2}\theta (\del_1\del'_2 -
\del_2\del'_1)}f(x)g(x')|_{x=x'} }
where the noncommutativity parameter $\theta_{ij}$, given in terms of
$B$ by\seiwit 
\eqn\bvalue{
\theta_{ij}= -(2\pi\alpha')^2 \Big({1\over g+2\pi\alpha' B}B
{1\over g-2\pi\alpha' B}\Big)_{ij}}
has the single component $\theta_{1\,2}=\theta$.

Let $V(T)$ denote the universal tachyon potential\senuniv\ on an
unstable D-brane. The action for such a non-BPS D-brane is given by
\refs{\senworld,\panda,\garou}:
\eqn\nbpseq{
 \mu^{(u)}_p \int d^{p+1}\sigma~V(T)\sqrt{-det(g_{\mu\nu}+
2\pi\alpha'{\cal F}_{\mu\nu} +  2\pi\alpha'\del_\mu T \del_\nu T)}}
where $g_{\mu\nu}$ is the closed string metric and $\cal F_{\mu\nu}$
is the linear combination of $F_{\mu\nu}$ and $B_{\mu\nu}$. We claim
that in the presence of a B-field, the tachyonic part of the D2-brane
worldvolume action is of the form
\eqn\dtwolag{
S =\int d^2x dt \left( \del^\mu T \del_\mu T - V(*T) \right)}
where the notation $V(*T)$ represents the universal tachyon potential
with all products replaced by $*$ products.

To argue this, note that the replacement of $V(T)$ by $V(*T)$
apparently violates universality of the tachyon potential. However,
precisely because this universality holds only in the zero-momentum
sector\senuniv, we can expect a violation of universality that
vanishes at zero-momentum. This is true of the $*$ product, which
reduces to the ordinary product on constant functions. Thus,
noncommutativity and universality of the tachyon potential together
amount to a derivation of the above action\foot{We are grateful to
K. Hori for discussions on this point.}.

For unstable D-branes in type II string theory, $V(T)$ is known to be
an even function of $T$ and is believed to have a double-well
shape. Let it take its minima at $T=\pm T_0$. At the minimum, the
negative potential energy localized on the brane should cancel the
unstable D2-brane tension $\muu2$, consistent with the decay of the
D2-brane into vacuum. This gives rise to the beautiful
equation\senreview:
\eqn\seneqn{
\muu2 + V(T=T_0)=0 }

Besides decay into the vacuum, the simplest decay mode of the unstable
D2-brane is via kink condensation in this double-well potential,
leading to a stable D-string. The D-string charge arises from the 
Chern-Simons coupling\refs{\hork,\senworld,\bcr}:
\eqn\cscoupling{
{1\over 2T_0} \int dT\wedge B_{RR} }
where $B_{RR}$ is the Ramond-Ramond 2-form in type IIB. For a kink
solution along say $x^2$, we have $\int dx^2 \del_2 T = 2T_0$ and we
are left with unit coupling to $B_{RR}$ and hence unit D-string
charge. An anti-kink would produce an anti-D-string.

We work in the limit of large $\theta$. It is possible to achieve this limit,
while keeping the open string metric\seiwit\ fixed, by a suitable choice
of scaling for $\alpha'$, $B$, and the closed string metric $g$, for example,
the limit
\eqn\scaling{\alpha'\to{\sqrt\epsilon},{\hskip10pt}
B\to{\sqrt\epsilon},{\hskip10pt} g\to\epsilon^2}
with $\epsilon\to 0$. Following Ref.\gms, it is 
convenient to scale the coordinates by $x^i \rightarrow \sqrt\theta
x^i$, and one ends up with the action
\eqn\scaledaction{
S =\int d^2x dt \left( \del^\mu T \del_\mu T - \theta V(*T) \right)}
where now the $*$ product is as in Eqn.\starprod\ but with $\theta=1$.

At infinite $\theta$, we can look for classical solutions by just
solving $\del V(*T)/\del T =0$. For a fairly arbitrary tachyon
potential (but in particular, without a linear term), an infinite set
of non-constant solitonic solutions of this equation was found in
Ref.\gms. In particular, one solution is provided by starting
with the function $\phi_0(x) \equiv 2 e^{-r^2}$ and writing
\eqn\gmsbasicsoln{
T(x) = \lambda_i \phi_0(x) }
where $\lambda_i$ is any nonzero minimum of the ordinary
potential. This works by virtue of the fact that (with $\theta=1$), we
have $\phi_0 * \phi_0 = \phi_0$.

Note that this soliton decays to zero at infinity. Although this was
not explicitly stated, the analysis and examples of Ref.\gms\ dealt
entirely with potentials having a quadratic minimum at the origin,
hence the soliton does in fact tend asymptotically to the vacuum. Our
case is just the opposite: a tachyonic potential has a quadratic
maximum at the origin. One can of course go from one to the other by
shifting $T(x)$ by a specific constant, though general shifts of
$T(x)$ are not allowed --- since they introduce a linear term in
$V(T)$, which in turn invalidates the solutions of Ref.\gms.  As we
will now see, a small modification of the example above by such a
constant shift will give rise to an interesting physical phenomenon
with unstable branes. Suppose we first take the classical solution:
\eqn\ourbasicsoln{
T(x) = T_0(1- \phi_0(x)) }
Note that, like $\phi_0(x)$, $(1-\phi_0(x))$ also squares to itself
under the $*$ product. However, it varies from $+1$ at spatial
infinity to $-1$ at the origin. Thus the configuration $T(x)$ above is 
a soliton that interpolates from $T_0$ at spatial infinity to $-T_0$
at the origin. One can of course take the negative of this solution
and it interpolates in the opposite way.

A key insight into the physical meaning of this process can be
obtained by looking at the total energy of this soliton. The energy
density of the soliton is easily evaluated using 
$(1-\phi_0) * (1-\phi_0) = (1-\phi_0)$:  
\eqn\solitonenergy{
V\big(T_0 (1-\phi_0(x)) \big) = \big(1-\phi_0(x)\big)V(T_0) }
Hence the total energy density on the D2-brane, with the
noncommutative soliton excited, is
\eqn\dtwoen{
\theta (\muu2 + \big(1-\phi_0(x) \big)V(T_0))}
Using Sen's conjecture for unstable brane decay, Eqn.\seneqn, this
works out to be $\theta\phi_0(x)\muu2$. This has to be interpreted as
the energy of the decay product.
We see that this decay product is an object whose energy is localised
very close to the origin, so it can be considered an exotic
0-brane. It is planar and its energy distribution falls off
exponentially with the distance away from the origin. Indeed, it is a
D0-brane in a sense, since a string that was ending on the original
D2-brane can continue to end on regions where there is finite D2-brane
tension.  Hence we interpret the noncommutative soliton above as
describing the decay of a planar unstable D2-brane into a localised
configuration at the origin. We will argue in the next section that
this configuration is really a smeared D-string.

It is noteworthy that the energy of the noncommutative soliton can be
calculated exactly without a detailed knowledge of the tachyon
potential. This is due to the fact, pointed out in Ref.\gms, that the
soliton solution depends on no details of the potential except
its value at the minimum. Remarkably, at the present stage of
understanding of string theory, this is the only thing about the
tachyon potential on D-branes that we {\it do} know reliably, thanks
to Eqn.\seneqn.

It is clearly important to analyse various properties of this D0-brane
including its stability (in the presence of large
noncommutativity). The above solution will turn out to be classically
unstable, as we will argue in Section 4.

There is, in fact, another classical solution which is most interesting from
the point of view of stability. This is given by:
\eqn\stablesolnzero{
T(r) = T_0(1-2\phi_0(r)) }
That these are classical solutions follows from the fact that they can
equivalently be written
\eqn\stableequiv{
T(r) = T_0(1-\phi_0(r)) + (-T_0)\phi_0(r) }
in which form it is evident that they are the
superposition\foot{Normally, one would not expect superpositions of solutions
to be solutions in a non-linear theory. However, the
orthogonality properties of the $\phi_n$ under the star product allow us to
superpose orthogonal solutions.}
of two noncommutative solitons, one of which asymptotes to 0 and the
other to $T_0$. Another way of seeing that these are solutions is that
although the function $(1-2\phi_0(r))$ does not square to itself under
the $*$ product (rather it squares to 1), its ${\it odd}$ powers are
all equal to itself, and that is sufficient because the tachyon
potential is even.

At the origin this solution tends to $-3 T_0$. We will see
later that this solution is metastable. The energy of this solution is
\eqn\stableenergy{
\eqalign{
V(T_0(1-2\phi_0(r))) &= (1-\phi_0(r))V(T_0) + \phi_0(r) V(-T_0)\cr
&= V(T_0)\cr }}
This is degenerate with the energy of the vacuum solution. Thus, in
this decay mode the original D2-brane has decayed into a nontrivial
configuration that has the same energy as the vacuum.

Before turning to the issue of stability, we look at some generalizations.
A complete set of radially symmetric solutions to the equation
$\phi * \phi = \phi$ was worked out in~\gms . They satisfy the relation:
\eqn\phiids{
\eqalign{
\phi_n(r) * \phi_m(r) &= \delta_{m\, n}\; \phi_n(r)\cr
\sum_{n=0}^{\infty} \phi_n(r) &= 1 \cr }}
These functions may be found as follows (the following constitutes a
quick alternate derivation to the one in Ref.\gms). Define the
generating function
\eqn\genfn{
\phi(r,z) = \sum_{n=0}^{\infty} \phi_n(r)z^n}
in terms of which the above equations amount to
\eqn\genfnprop{
\eqalign{
\phi(r,z) * \phi(r,z') &= \phi(r,zz')\cr
\phi(r,z=1) = 1\cr}}
The first of these equations may be solved by going to Fourier space
in $x$ (recall that above, $r=|x|$). One finds that, with a radially
symmetric ansatz, the solution is Gaussian in Fourier space and hence
also in coordinate space. Inserting this ansatz and solving the above
equations, one ends up with the generating function:
\eqn\genfnsoln{
\phi(r,z) = {2\over 1+z} e^{ {z-1\over z+1} r^2 }}
Expanding in powers of $z$ one finds:
\eqn\phinsoln{
\phi_n(r) = 2 (-1)^n e^{-r^2} L_n(2 r^2)}
where $L_n$ are the Laguerre polynomials, in agreement with the
solution in Ref.\gms.

We can now write the obvious generalization of Eqn.\ourbasicsoln ,
$T_n(r) = T_0(1-\phi_n(r))$, noting that $(1-\phi_n(r))$ squares to itself
under the star product. This soliton is also a D0-brane, but is $\sqrt{n}$
times larger than the original one. While for even $n$ the solution
interpolates between one vacuum $T=T_0$ at infinity to the other
vacuum $T=-T_0$ at the origin, for odd $n$ the situation is quite
different. From $\phi_n(r=0) = 2(-1)^n$, we see that the solutions for 
odd $n$ interpolate between the vacuum $T=T_0$ at infinity and a
non-vacuum configuration $T=3T_0$ at the origin. There is, of course,
no requirement that the soliton should go to a vacuum at the origin.

It is also easy to see that $T(r) = T_0(1-f(r))$, with
\eqn\fofr{
f(r)=\sum_{i\in I}\phi_i(r) }
where the sum runs over a finite index set $I$ of the $\phi_i$'s, each
counted exactly once, is also a classical solution, since such a
function $f(r)$ also squares to itself under the star
product. Functions of the form $(1-f(r))$ with $f(r)$ as above are, in
fact, the most general functions that square to themselves and have
the asymptotic behaviour of a solitonic solution for a tachyonic
potential, but as we will see in a moment, these are not the most
general extrema of the noncommutative tachyon potential.

It is interesting to note the relation
\eqn\infsum{(1-\phi_n)=\sum_{m\neq n} \phi_m}
so that the above solitons may be regarded as the superposition of an
infinite set of the noncommutative solitons of \gms. Note that since
$\phi_n$ vanishes at infinity, the asymptotics of the functions
$(1-\phi_n)$ and $\phi_n$ are very different. There is, however, no
contradiction with \infsum\ because the RHS is an infinite sum.

It is also easy to write down a generalization of Eqn.\stablesolnzero.
Given that the tachyon potential is an even function, we can find a
class of solutions with zero total energy using functions $f(r)$ that
satisfy $f(r)*f(r)=1$. Using any such function we can define a 
solution $T(r)=T_0 f(r)$ that has $V(T(r)) = V(T_0)$ and hence
vanishing total energy.

Expanding such functions over the $\phi_n(r)$ we have:
\eqn\expandfns{
f(r) = (1 + \sum_{i\in I}\lambda_i \phi_i(r) ) }
where again $I$ is a finite index set of distinct elements. One sees
that
\eqn\fnsolns{
f(r)* f(r) = 1 + \sum_{i\in I} \lambda_i(\lambda_i + 2)\phi_i(r) }
from which it is clear that each $\lambda$ must be equal to 0 or $-2$.
Thus the classical solutions 
\eqn\genstable{
T(r) = T_0 (1 -2 \sum_{i\in I}\phi_i(r) ) }
all have zero energy and will turn out to be metastable.

Finally, we can superpose the two kinds of solutions above to obtain the
most general solitonic solution (upto an overall choice of sign, of course)
for a double well potential 
\eqn\gensoliton{
T_r = T_0(1 - \sum_{i\in I} \phi_i) + (-T_0) (\sum_{j\in J} \phi_j)}
where $I$ and $J$ are finite index sets with $J\subset I$. This has
positive energy when $J$ is a proper subset of $I$, reducing to \fofr\ when
$J$ is the empty set. When $J=I$, on the other hand it reduces to
\genstable, and has zero total energy. 

More precisely, the total energy of such solitons does not exactly
vanish, but for any finite $\theta$, it is suppressed relative to
normal D-brane energies by a factor of ${1\over\theta}$. At finite
$\theta$, kinetic terms would contribute a small mass correction, as a
result of which the energy of these solitons (following analogous
arguments in Ref.\gms) is generically $V(T_0)+{\cal
O}({1\over\theta})$. When we rescale coordinates to absorb a
$\sqrt\theta$ in them, the total energy has a factor $\theta$
multiplying it, as in Eqn.\dtwoen. In these coordinates the mass of
the above solitons is therefore finite, though the tensions of the
standard stable and unstable D-branes of the theory are proportional
to $\theta$. If we revert to the original coordinates then the
standard D-branes have finite tension and the solitons described above
have total energies of order $1\over\theta$.

The extension of the results in this section to unstable Dp-branes for
various $p>2$ (odd in type IIA and even in type IIB) is
straightforward, for the case of noncommutativity only along two
spatial directions. More interesting extensions to higher branes will
be discussed in a subsequent section.

\newsec{Noncommutative Soliton as a Smeared D-string}

In this section we consider the D-brane charge associated with the
above solitonic branes. Although the noncommutative solitons are not
topological, and hence the exotic 0-branes of the previous section
cannot carry a global RR charge, we can gain some insight into their
nature by looking at what RR charge they carry locally.

As described for one special case in Eqn.\cscoupling\ above, there is
a Chern-Simons coupling on the non-BPS 
$p$-branes\refs{\hork,\senworld,\bcr}
\eqn\csterm{{1\over 2 T_0} \int dT\wedge C_p}
where $C_p$ is the Ramond-Ramond $p$-form, and $T$ is the tachyon. In
the limit when the noncommutativity parameter
$\theta\rightarrow\infty$, we need only keep terms in the action which
scale like $\theta$ (after coordinate redefinition). However, the
Chern-Simons terms above, being topological, do not pick up any such
factor, so in the $\theta\rightarrow\infty$ limit, they might appear
to be irrelevant. This, however, is not the case. When $T$ is a kink
along one of the world-volume directions, we expect the solution to
correspond to a $(p-1)$-brane, and therefore there must exist a
coupling $\int C_p$ on its world-volume. The only source for such a
coupling is the Chern-Simons coupling above, which can therefore not
be neglected, even in the limit $\theta\rightarrow\infty$.

Now consider, as before, the unstable Type IIB D2-brane in a
background $B_{NS}$ field, and let us consider the tachyon background
$T=T_0(1-f)$, where $T_0$ is the minimum of the tachyon potential, and
$f$ is, say, the Gaussian solution $\phi_0$. 
Since $T_0$ is constant, \csterm\ yields
\eqn\cstwo{- {1\over 2 T_0} \int T_0\, d\phi_0\wedge B = 
- {1\over 2} \int \phi_0'(r)\, B_{t\, \omega}\, 
dr\wedge dt\wedge d\omega}
where $B$ is the Ramond-Ramond $2$-form, so that the soliton locally
carries D-string charge. In fact, since, $\int \phi_0'(r)\, dr = -
\phi_0(0) = - 2$, we get precisely $\int B_{t\, \omega}\, dt\wedge d\omega$,
implying that the soliton carries unit D-string charge, and
can in fact be interpreted as a D-string winding along the angular
coordinate\foot{We will use $\omega$ to refer to the angular variable
of the polar coordinates, reserving the more conventional $\theta$ for
the noncommutativity parameter.} $\omega$. The function $f(r)$ can be
interpreted as the distribution function for the D-string, so that our
soliton describes a D-string smeared over the radial coordinate $r$,
winding around the origin. 

Such a D-string would not normally be stable, but would collapse by
contracting to a point at the origin. This is equivalent to saying
there is no global RR charge carried by this configuration. However,
the above argument indicates that there is a local RR charge and
enables an interpretation of the exotic decay product in terms of
D-strings. To the extent that noncommutativity stabilizes some of the
exotic 0-branes, which is discussed in a subsequent section, it will
be equally true that such a smeared, winding D-string is stabilized.

Note that if we choose $T=-T_0(1-\phi_0)$, we get an anti-D-string
instead. It is also easy to verify that, if we choose $f = \phi_n$,
the soliton carries D-string charge $(-1)^n$.  Again, if $f =
\phi_0 + \phi_1$ (so that $T= T_0(1-f)$ is still a classical solution)
the total D-string charge vanishes. However, because the two
functions $\phi_0$ and $\phi_1$ do not add to zero, we should really
interpret this soliton as a dipole of D-strings smeared over the
$(r,\omega)$ plane. This argument extends to the situation when we
have $f = \sum \phi_i$ for some finite set of $\phi_i$'s.

Finally, for the solutions that are degenerate with the vacuum, for
example $T(r) = T_0(1-2 \phi_n(r))$, we find 2 units of D-string or
anti D-string charge locally, depending on whether $n$ is even or odd.
For $T(r) = T_0(1-2 \phi_0(r) -2\phi_1(r))$, we again obtain a dipole
of D-strings. The extension to solitons of the form \genstable\ and
\gensoliton\ is obvious.
 
\newsec{Stability of the Decay Products}

Among the various classical solutions to the noncommutative tachyon
theory, we found the class of solutions $T(r) = T_0(1-2\sum_{i\in
I}\phi_i(r))$ with energy $V(T_0)$. This energy is negative, and is in
fact just the same as we would have obtained by choosing the trivial
classical solution (valid in both commutative and noncommutative
cases) $T(r)=T_0$. Adding it to the tension $\muu2$ of the unstable
D2-brane, we get a total energy 0. Thus (in the limit of infinite
noncommutativity) we find that an unstable D2-brane can decay into a
configuration that, while quite different from the vacuum\foot{The
key difference from the vacuum is that this configuration carries
local D-string charge, as described in the previous section.}, is
nevertheless degenerate with it. In this situation one would expect
the decay product to be stable.

One should, however, think of this as an approximation. As discussed
in Section 2, there will be corrections due to finiteness of the
noncommutativity parameter $\theta$. These corrections will tend to
raise the energy of the decay product slightly above the vacuum, in
which case it will be metastable. More information about the shape of
the tachyon potential in string theory than is presently available
would be needed to estimate the lifetime of this state.

The other solutions we described, such as the solution
$T(r)=T_0(1-\phi_n(r))$, are instead classically unstable. The easiest
way to see this is to rewrite the total energy as follows\foot{We wish
to thank Rajesh Gopakumar for the following argument.}
\eqn\newpot{
V(T) + \muu2 = {\widetilde V}(T')}
with $T'=T-T_0$. Then, ${\widetilde V}$ has degenerate minima at $0$ and
$-2T_0$, and a maximum at $-T_0$, and belongs to the general class of
potentials discussed in \gms. From this, it is easy to see that the solution
$T(r)=T_0(1-\phi_n(r))$ corresponds to $T'(r) = - T_0\phi_n(r)$, which is
unstable, as its energy can be decreased by scaling by a constant near unity.
In other words, the energy of $T(r) = T_0(1-\phi_n(r)) + \epsilon\phi_n(r)$, 
(for small $\epsilon$) is {\it lower} than $T(r)$ by ${\cal O}(\epsilon^2)$.

These solutions have a finite energy above the vacuum and will decay
into it classically. Hence they should be thought of as describing the
decay of an unstable D2-brane into a kind of unstable D0-brane. The
situation is somewhat reminiscent of the case of a brane-antibrane
pair, which can decay into an unstable brane of one lower dimension by
condensing an unstable tachyonic kink.

It is fascinating that, from the discussion in the previous section,
the classically stable solitons all carry even D-string charge, while
all the solitons with odd D-string charge are unstable. Note of
course, that not all even charge solitons are stable. For example, the
dipole $T_0(1 - \phi_0 - \phi_1)$, and the charge 2 soliton $T_0(1 -
\phi_0 - \phi_2)$ are classically unstable.

\newsec{Noncommutative Decays of Higher $p$-branes}

We can generalize the argument of Section 2 to higher brane solitons.
Consider an unstable D3 brane of Type IIA, with worldvolume along
$(x^0,x^1,x^2,x^3)$ in the presence of a constant background
Neveu-Schwarz B-field along $(x^1,x^2)$ directions. We expect the
geometry along the $(x^1,x^2)$ plane to become noncommutative, while
$(x^0,x^3)$ remain commutative. We can now take the limit of large
noncommutativity parameter $\theta = \theta_{1\,2}$, and perform the
appropriate rescalings of the coordinates. Now consider the tachyon
background $T=K(x^3)(1-f(x^1,x^2))$, where $K(x^3)$ is the kink along
the commutative direction $x_3$, and $f$ is one of the noncommutative
solitons $\phi_n$. Then from the coupling
\eqn\csthree{{1\over 2 T_0} \int dT \wedge C}
where $C$ is the Ramond-Ramond 3-form, we get 
\eqn\dtwocharge{{1\over 2 T_0} \int_{D3} (\, (1-f)\, dK  - K\, df\, )\wedge C
=  \int_{D2} (1-f)\, C - {1\over 2 T_0} \int_{D3} K\, df\wedge C}
where the first term on the RHS is an integral over the D2-brane obtained
from the kink $K$. In the absence of $f$, we should get therefore get a D2
brane. When $f=\phi_n$, however, from the term $\int_{D2} C (1-f)$ it appears
that the D2 brane charge is altered by an ``irrational'' amount, since
$\int_{D2} f(r)\, dx^1 dx^2 = 2\pi$, leading to a apparent contradiction.
However, we should really examine the charge density, that is, the charge per
unit D2-brane volume, and recall that the volume of the D2 brane is infinite.
Thus the 1 in $(1-f)$ still contributes 1 unit of D-brane charge but since
$\int_{D2} f(r)\, dx^1 dx^2$ is finite, it does not affect the charge
density, so it can be ignored.

We also have a second term in \dtwocharge\ :
\eqn\dipole{- {1\over 2 T_0} \int K\, df\wedge C} 
As in the D2 brane case above, this gives
\eqn\dipolecharge{{1\over T_0} \int K(x^3)\, C_{0\, \omega 3}
dx^0\wedge d\omega\wedge dx^3} 
Since the kink $K$ goes from $-T_0$ to $T_0$ as $x^3$ goes from
$-\infty$ to $\infty$, we see that we get zero total D2 brane charge.
However, we really have a dipole of D2 branes, with the anti-brane
at $x^3 < 0$ (and the radial coordinate
$r=\sqrt{(x^1)^2+(x^2)^2}$ ) and the brane at $x^3 > 0$
(and $r$).

Finally, from the coupling in the non-BPS D3 brane action:
\eqn\irrational{{1\over 2 T_0} \int dT \wedge A \wedge B_{NS}} 
where $A$ is the Ramond-Ramond 1-form, and $B_{NS}$ is the background
Neveu-Schwarz B-field, we might expect the solitons to carry D0-brane charge. 
However, since the Neveu-Schwarz background is not quantized, we cannot
in general expect integer (or even rational) D-brane charge from \irrational .
We expect the resolution to this puzzle to be along the lines of \wati ,
with the D-brane charge from \irrational\ being cancelled by bulk terms.

The above exercise can be repeated with the metastable solutions of
the type $K(1-2\phi_n(r))$ along the noncommutative directions, with similar
conclusions.

\newsec{Conclusions}

It is pleasing that the field-theoretic study of noncommutative
solitons initiated in Ref.\gms\ has such an elegant application to the
scalar field theory of tachyons on the world-volume of unstable
D-branes in superstring theory.  A more detailed understanding of the
physical significance of this decay process, of the stability of the
end products, and of modifications due to finite rather than infinite
$\theta$ is clearly desirable.

In particular, consider the smeared D-string configurations that
describe metastable decay products. It would be desirable to explain,
from the spacetime (as opposed to brane worldvolume) point of view, why
such configurations are rendered metastable by a suitable constant
B-field\foot{Steve Gubser has suggested that these configurations may
be spinning D-strings stabilised by their centrifugal force.}.

It would be interesting to extend our results to Dp-branes of other
dimensions, and to the noncommutative gauge theory that also resides
on unstable D-branes. One could also consider the complex tachyon on a
brane-antibrane pair, which couples to a worldvolume gauge field,
unlike the real noncommutative tachyon studied in this paper. In such
a case, one might combine noncommutative solitons with orthogonal
tachyonic vortices in the noncommutative scalar-gauge theory that
resides on brane-antibrane pairs. Finally, the physical effects of
noncommutativity along, rather than transverse to, a topologically
stable kink or vortex remain to be explored.
\bigskip

\noindent{\bf Acknowledgements:}

We would like to thank Ori Ganor, Rajesh Gopakumar, Steve Gubser, Jeff
Harvey, Joydeep Majumdar, Shiraz Minwalla, Nemani Suryanarayana and
Sandip Trivedi for helpful discussions. We are particularly grateful
to Rajesh Gopakumar for reading and helping us correct a preliminary
version of the manuscript. S.M. would like to thank Nati Seiberg and
the Institute for Advanced Study, Princeton, where this work was
initiated, for hospitality. The work of K.D. was supported by DOE grant
No. DE-FG02-90ER40542, and that of G.R. by NSF grant
No. NSF-DMS-9627351.

\vfill\eject

\listrefs
\bye